\documentclass[11pt]{article} 
\usepackage{amsmath}

\usepackage[utf8]{inputenc} 

\usepackage{geometry} 
\usepackage{graphicx} 

\usepackage{booktabs} 
\usepackage{array} 
\usepackage{paralist} 
\usepackage{verbatim} 
\usepackage{subfig} 

\usepackage{fancyhdr} 
\usepackage{sectsty}
\allsectionsfont{\sffamily\mdseries\upshape} 

\usepackage[nottoc,notlof,notlot]{tocbibind} 
\usepackage[titles,subfigure]{tocloft} 

\title{\bf{On Hamilton's Principle for Discrete and Continuous Media}}
\author{\bf{ Vassilios K Kalpakides}\\ 
Department of Materials Science and Engineering,\\
University of Ioannina, Ioannina, GR-45110, Greece\\
\\
\bf{Antonios Charalambopoulos}\\
School of Applied Mathematics and Physics\\
National Technical University of Athens, GR-15780, Zografou, Greece}

\date{} 

\begin{document}
\maketitle
\begin{abstract}
In an attempt to generalize the Hamilton's principle, an action functional is proposed which, unlike the standard version of the principle, accounts properly for all initial data and the possible presence of dissipation. To this end, the convolution is used instead of the $L^2$ inner product so as to eliminate the undesirable end temporal condition of Hamilton’s principle. Also, fractional derivatives are used to account for dissipation and the Dirac delta function is exploited so as the initial velocity to be inherently set into the variational setting. The proposed approach applies in both finite and infinite dimensional systems.\\
\\
{\bf Keywords:} variational principle, initial value problem, convolution,fractional derivative, dissipation
\end{abstract}

\section{Intorduction}
Hamilton's principle is usually considered the proper variational setting for a dynamical system. It is based on the celebrated {\it  functional of action} which, as concerns time, is defined on an interval $[t_1,t_2]$, where  $t_1$  and  $t_2$ denote the initial and the  final time  of the motion. Moreover, one needs data for the state of the system at both moments $t_1$ and $t_2$ so that the domain of action's  functional can be defined \cite{BEDFORD}. The extremum of the action's functional is connected, through a well--known procedure, with the Euler--Lagrange equations, i.e., the field equations  of the dynamical system.  However, there is a contradiction between the domain of the action functional and the corresponding  initial value problem because the latter admits temporal conditions only at the  initial moment $t_1$. The status of the system at the final time $t_2$ is not only unknown but it is  what one is looking for, together with the behavior of the system for any finite $t>t_1$.  As Tonti remarked \cite{TONTI1973}, this deficiency comes from the use of the $L^2$ inner product. For instance, consider the second order diferrential operator on $[t_1,t_2]$ which  is formally symmetric with respect to $L^2$, while it fails to be symmetric when initial conditions are taken into account. Roughly speaking, after integrating by parts  twice, the corresponding bilinear form will display values of the involved function and of its derivative at the end point $t_2$ for which no data are available. Summing up,  Hamillton's principle may derive the proper field equation, but it fails to produce the corresponding initial conditions. \\
\\
A second  deficiency of the Hamilton's principle will be revealed when one negotiates with a non-conservative system, where first order derivative  be involved in the differential equation. For instance, in cases where friction or any kind of inelastic behavior is  present, the standard Hamilton's principle fails, not only as concerns the initial conditions, but to derive the field equation as well. Various methods to generalize the principle so as to cover the dissipative phenomena have been proposed as referred in \cite{BEDFORD,TONTI1973,DARGUSH-et-al2012}. Nevertheless, none of them are so much rigorous, simple and elegant as the standard version of the Hamilton's principle for conservative systems.\\
\\
Another discrepancy appears while comparing the extremum  principles for dynamics and statics of continuous media. 
Hamilton's principle for elasticity is usually regarded as a generalization to dynamics of the extremum principle for  the total potential energy which concerns the equilibrium of an elastic body.  For the latter, one can find  a rigorous mathematical justification in any standard textbook of applied functional analysis, for instance, see the theorem for the minimum of a quadratic functional in \cite{ZEIDLER}.  On the contrary, there is not any satisfactory mathematical justification for the former, i.e., for Hamilton's principle.\\
\\
From the aforementioned drawbacks, it becomes obvious that  Hamilton's principle must be reconsidered so as to meet the proper initial conditions, to account for dissipation and to be set  into a sound mathematical context.
Such an attempt has been initiated  by   Gurtin \cite{GURTIN1963,GURTIN1964,GURTIN1964b} who exploits the convolution to transform an initial--boundary value problem into a boundary value problem for which a variational principle can be formulated. Actually, the use of the convolution  instead of  $L^2$ inner product results in vanishing the undesirable terms which come out of the integration by parts. This idea has been proved fruitful and  has been used and further developed by many researchers \cite{REDDY1976,REDDY1976b,RAFAL1968,LUO-et-al1988,PENG-et-al1996} to formulate variational principles sufficient to derive not only the field equations but the appropriate initial conditions as well. \\
\\
As concerns the non--conservative systems, a decisive step has been done  in 1996 by F. Riewe who proposed the use of fractional derivative in the action functional \cite{RIEWE1996,RIEWE1997}. Remarking that if the Lagrangian function contains derivative of the $n$--th order, the Euler--Lagrange equation will be of the $2n$--th order, he assumed that a first order derivative in the field equation -- for instance a friction force -- demands a Lagrangian function containing  a term of half order derivative. This simple idea has been successfully used by a great number of authors to formulate variational principles for particle and continuum mechanics \cite{APOSTOL-et-al2013,APOSTOL-et-al2013b,DARGUSH2012,DARGUSH-et-al2012,DARGUSH-et-al2015,DARGUSH-et-al2016,DARRAL-et-al2018,KIM2014,KIM2017,GARRA-et-al2018,KARAMANY-et-al2011,MALIN-et-al2012}.\\
\\
An efficient variational principle, the so--called {\it mixed convolved action principle}, has been introduced by Dargush and his co-workers \cite{DARGUSH2012,DARGUSH-et-al2012,DARGUSH-et-al2015,DARGUSH-et-al2016} to remedy all the above mentioned discrepancies of Hamilton's principle. It is  a variational principle  of mixed type, where  the action functional is defined in terms of both the displacment and the impulse of stress. Taking the variation of the action functional with respect to these variables, one obtains two equations: the field equation and the constitutive relation.\\
\\
However, making use of the stress  impulse  (or of the force impulse), one introduces some ambiguities in relation with the proper integration interval. If one selects the positive semi-axis as integration interval, that is, if one defines the force impulse as 
$$
J(t)=\int_0^t{f(s)ds},\quad t\in [0,\infty ),
$$
then $J(0)=0$ and the corresponding initial condition will become necessarily homogeneous, an undesired restriction for the initial data.
Alternatively, the impulse is often defined as 
$$
J(t)=\int_{-\infty}^t{f(s)ds},\quad t\in (-\infty,+\infty)
$$
so as the  history of the loading to provide the necessary information at the onset of the motion. Though this approach impoves the behavior of the impulse at the initial time of the previous definition, it is unnecessarily extended to the entire past history to catch data concerning only the time $t=0$.\\
\\
The present work  aims at a systematic derivation of variational principles formulated  in terms of the diplsacement field. That means  the constitutive relations are a priori assumed,  and so are  incorporated into the field equations. More specifically,  our goal is first to define a proper functional and its domain which will be the  set of the kinematically admissible (with regard time or space--time) displacements.  Then,  to prove that  the variation of this functional with respect to the displacement field  provides the desired initial or initial--boundary value problem. Regarding the continuous systems, the above procedure will apply in dynamics of bodies with linear elastic and viscoelastic response. \\
\\
Of central importance in the present  approach is the consideration that the external excitation consists of two parts. The first one is a classical continuous function interpreting the impact of the environment on the system during the evolution. The second part is a generalized function of Dirac type which interprets a force acting on  the system only at the initial time, providing the proper linear momentum just at    the onset of the evolution. This proves to be necessary so as the initial condtiion for the velocity to be built into the variational principle. Otherwise, one has to impose it externally.  \\
\\
 A consequence of our analysis is that  the initial conditions,  in some way, will  be set on equal footing with the boundary conditions. Consider  the variational principle of elastostatics, the Dirichlet boundary conditions find theirselves in the domain of the functional, i.e., in the set of the  kinematically admissible functions. On the other hand, the Neumann boundary conditions resides in a term of the total potential energy functional. In an analogous manner, it will be shown that the initial displacement contributes to the definition of the domain of the action functional, while the initial  velocity field contributes through a novel term to the action functional.\\
\\
 In Section 2, some necessary mathematical preliminaries concerning fundamental properties of the convolution, the fractional derivative and the fundamental lemma of variational calculus are provided in order to be used in the following sections. In Section 3,  generalized Hamilton's principles for dynamical systems of a single degree of freedom for conservative and dessipative systems are formulated and then proved. In Section 4, variational principles for dynamics of one--dimensional elastic and viscoelastic bodies are presented and finally in Section 5, some concluding remarks are discussed.

\section{Preliminaries}
In this section some definitions and existed results that are necessary to the subsequent analysis are presented for the sake of completeness of the paper and to familiarize the interested reader with the used denotation. 
\subsection{The convolution}
Let the functions $f$ and $g$ are continuous on $[0,\infty)$. The convolution of $f$ and $g$  is the function $[f,g](t)$ on $[0,\infty)$, defined by
\begin{equation}
[f,g](t)=\int_0^tf(s)g(t-s)ds.
\end{equation}
If $f$ and $g$ are functions of space and time, i.e,  $ f=f({\bf x},t),\quad g=g({\bf x},t),\quad {\bf x}\in \Omega\subset {\bf R}^3$ the temporal convolution will be defined as
\begin{equation}
[f,g]({\bf x},t)=\int_0^tf({\bf x},s)g({\bf x},t-s)ds,
\end {equation}
where for any ${\bf x}\in\Omega$, $ f({\bf x},\cdot )$ and  $ g({\bf x},\cdot )$  are  continuous functions on $[0,\infty )$. 

One can confirm that the convolution fulfills the following properties \cite{GURTIN1964}
\begin{eqnarray}
&&[f,g]=[g,f],\\
&& [f+g,h]=[f,h]+[g,h],\\
&& [f,g]=0 \hbox{ implies either }f=0  \hbox{ or } g=0.
\end{eqnarray}
\subsection{The integration by parts}
Very often, one needs to use a convolution of the form
\begin{equation}
\left[ v,u'\right](t)=\int_0^tv(s)u'(t-s)ds,
\end{equation}
where 
$$
u'(t-s)=\frac{du}{dz}(z),\quad  \hbox{for}\quad  z=t-s.
$$
Defining 
$$
w(s):=u(t-s),
$$
one can write
\begin{equation}
w'(s)=\frac{du}{dz}\frac{dz}{ds}=-u'(z)=-u'(t-s).
\end{equation}
Then eq. (6) with the aid of eq. (7) becomes
\begin{eqnarray}
\left[ v,u'\right](t) &=&\int_0^tv(s)u'(t-s)ds=-\int_0^tv(s)w'(s)ds=\left[-v(s)w(s) \right]_0^t+\int_0^tv'(s)w(s)ds\nonumber\\
&=&\left[-v(s)u(t-s) \right]_0^t+\int_0^tv'(s)u(t-s)ds\nonumber\\
&=&\left[-v(t)u(0)+v(0)u(t) \right]+\int_0^tv'(s)u(t-s)ds\nonumber\Rightarrow\\
\left[ v,u'\right](t)& =&\left[v(0)u(t) -v(t)u(0)\right]+\left[ v',u\right](t).
\end{eqnarray}\\
\\
If $v$ and $u$ were functions of space and time, then one could  write
\begin{equation}
\left[ v,\frac{\partial u}{\partial s}\right]({\bf x},t)=\int_0^tv({\bf x},s)u_{t-s}'({\bf x},t-s)ds,
\end{equation}
where it is understood that
$$
u'_{t-s}({\bf x},t-s):=\frac{\partial u}{\partial z}({\bf x},z),\quad z=t-s.
$$
Furthermore, if 
$$
u'_s({\bf x},t-s)=\frac{\partial w}{\partial s}({\bf x},s),\quad w({\bf x},s):=u({\bf x},t-s),
$$
then it holds
\begin{equation}
u'_{t-s}({\bf x},t-s)=-u'_s({\bf x},t-s).
\end{equation}
After that, one can write with the aid of eq. (8)
\begin{equation}
\left[ v,\frac{\partial u}{\partial s}\right]({\bf x},t) =\left[v({\bf x},0)u({\bf x},t) -v({\bf x},t)u({\bf x},0)\right]+\left[ \frac{\partial v}{\partial s},u\right]({\bf x},t)
\end{equation}
and
\begin{equation}
\left[ \frac{\partial v}{\partial s},\frac{\partial u}{\partial s}\right]({\bf x},t) =\left[\frac{\partial v}{\partial s}({\bf x},0)u({\bf x},t) -\frac{\partial v}{\partial s}({\bf x},t)u({\bf x},0)\right]+\left[ \frac{{\partial}^2 v}{\partial s},u\right]({\bf x},t).
\end{equation}
\subsection{The first variation of $[u',u']$}
For an arbitrary fixed $t<\infty$, consider the functional
\begin{equation}
J[u]=[u,u],\quad u\in {\cal D},
\end{equation}
where 
$$
{\cal D}=\{\phi \in C^1[0,t]\ |\ \phi(0)=u_0\}.
$$
The first variation of $J$ at $u$ with respect to $\eta$ is defined as 
\begin{equation}
DJ(u;\eta)=\frac{d}{d\epsilon}J[\tilde{u}]|_{\epsilon=0},
\end{equation}
where 
$\tilde{u}=u+\epsilon \eta$  is an arbitrary element of  $\cal D$,  provided that $\epsilon$ is a real parameter and $\eta \in {\cal D}_0$ with
$$
{\cal D}_0=\{\phi \in C^1[0,t]\ |\ \phi(0)=0\}.
$$
It is a matter of a simple calculation to confirm that 
\begin{equation}
\frac{d}{d\epsilon}J[\tilde{u}]|_{\epsilon=0}=\int_0^t[u(s)\eta (s)+\eta (s) u(s)]ds=[u,\eta]+[\eta,u]=2[u,\eta].
\end{equation}

Now consider the functional
\begin{equation}
J_1[u]=[u',u'],\quad u\in {\cal D}.
\end{equation}
Combining eqs (15), (8), invoking the fact that $\eta \in {\cal D}_0$ and  assuming in addition that $u''$ exists, one obtains
\begin{eqnarray}
\frac{d}{d\epsilon}J_1[\tilde{u}]|_{\epsilon=0}=2[u',{\eta}']&=&2\left(u'(0)\eta(t) -u'(t)\eta(0)+[u'',\eta]\right) \nonumber\\
&=&2\left(u'(0)\eta(t)+[u'',\eta]\right)
\end{eqnarray}
or
\begin{equation}
D[u',u'](u;\eta)=2\left(u'(0)\eta(t)+[u'',\eta]\right).
\end{equation}
\subsection{The first variation of $\int_{\Omega}[\frac{\partial u}{\partial s},\frac{\partial u}{\partial s}]$ }
Let the function $u=u({\bf x},s), {\bf x}\in \Omega, s\in [0,t]$ is sufficiently smooth so as, for any finite $t$, the functional 
$$
J[u]=\int_{\Omega}\left[\frac{\partial u}{\partial s},\frac{\partial u}{\partial s}\right]d{\bf x}
$$
exists for all  $u\in {\cal D}$, where
$$
{\cal D}=\left\{\phi \in C^1\left(\bar{\Omega} \times [0,t]\right)|\phi ({\bf x},s)=\hat{u}({\bf x},s) \hbox{ for all } {\bf x}\in \partial {\Omega}_D \hbox{ and }  \phi({\bf x},0)=u_0({\bf x})\right\}, 
$$
$\partial {\Omega}_D\subset \partial{\Omega}$ and $\hat{u}$,  $u_0$ are given functions.\\
For 
$$
\eta\in {\cal D}_0=\left\{\phi \in C^1\left(\bar{\Omega} \times [0,t]\right) | \phi ( \partial {\Omega}_D,s)=0  \hbox{ and }  \phi({\bf x},0)=0\right\}
$$ 
the first variation of $J$ at $u$ with respect to $\eta$  can be written
\begin{eqnarray}
DJ(u;\eta)&=&\frac{d}{d\epsilon}J[u+\epsilon\eta]|_{\epsilon=0}\nonumber\\
&=&\frac{d}{d\epsilon}\int_{\Omega}\int_0^t\left(u'_s({\bf x},s)+\epsilon{\eta}'_s({\bf x},s) \right)\left(u'_{t-s}({\bf x},t-s)+\epsilon{\eta}'_{t-s}({\bf x},t-s) \right)dsd{\bf x}|_{\epsilon=0}\nonumber\\
&=&\int_{\Omega}\int_0^tu'_s({\bf x},s){\eta}'_{t-s}({\bf x},t-s)+ {\eta}'_{s}({\bf x},s) u'_{t-s}({\bf x},t-s)dsd{\bf x} \nonumber\\
&=&\int_{\Omega}\left(\left[ \frac{\partial u}{\partial s},  \frac{\partial \eta}{\partial s} \right]+\left[ \frac{\partial \eta}{\partial s},  \frac{\partial u}{\partial s} \right]\right)d{\bf x},\nonumber
\end{eqnarray}
or
\begin{equation}
 DJ(u;\eta)=2\int_{\Omega}\left[ \frac{\partial u}{\partial s},  \frac{\partial \eta}{\partial s} \right]d{\bf x}.
\end{equation}
If $u$ is a litlle bit smoother \footnote{Actually, if $\partial^2 u/\partial s^2$ exists.}, the R.H.S of eq. (19) can be integrated py parts  according to eq.  (12)
$$
 DJ(u;\eta)=2\int_{\Omega}\left(\left[\frac{\partial u}{\partial s}({\bf x},0)\eta ({\bf x},t) -\frac{\partial u}{\partial s}({\bf x},t)\eta({\bf x},0)\right]+\left[ \frac{{\partial}^2 u}{\partial s},\eta\right]\right)d{\bf x},
$$
which becomes
\begin{equation}
 DJ(u;\eta)=2\int_{\Omega}\left(\left(\frac{\partial u}{\partial s}({\bf x},0)\eta ({\bf x},t) \right)+\left[ \frac{{\partial}^2 u}{\partial s},\eta\right]\right)d{\bf x}.
\end{equation}
\subsection{The fractional derivative  of order $1/2$}
The last years a vast literature concerning the application of fractional calculus in mechanics has been developed, see for instance  \cite{ATANASCOVIC2002,ATANASCOVIC-et-al2002,KARAMANY-et-al2011,GARRA-et-al2018,KATSIK2015,MALIN-et-al2012,TARASOV-et-all2015} to mention randomly just a few of them.  Here, we shall confine ourselves only to the  fractional derivative of order $1/2$, which we are going to use  in the next sections. In this subsection,  following mainly Dargush and Kim \cite{DARGUSH-et-al2012}, we simply provide some fundamental definitions and furmulae concerning the integration by parts and the variation of a functional containing derivatives of order $1/2$. For an analytical account of the subject,  we refer to \cite{OLDHAM,GORENFLO}. \\
\\
Let $u$ a sufficiently smooth function defined on the interval $[0,\tau]$. The( left)  Riemann-Liouville  integral and derivative of $u$  of order $1/2$ are defined as 
\begin{equation}
\left({\cal J}^{1/2}_{0+}u\right) (\tau)=\frac{1}{\Gamma (1/2)}\int_0^{\tau}\frac{u(s)}{(\tau -s)^{1/2}}ds,\quad\tau>0,
\end{equation}
\begin{equation}
\left({\cal D}^{1/2}_{0+}u\right)(\tau)=\frac{1}{\Gamma (1/2)}\frac{d}{d\tau}\int_0^{\tau}\frac{u(s)}{(\tau -s)^{1/2}}ds,\quad\tau> 0,
\end{equation}
respectively, where $\Gamma$ denotes the Gamma function.\\
Using the convolution, the following  integration by parts formula can be obtained  \cite{DARGUSH-et-al2012}:
\begin{eqnarray}
\int_0^tf(s)\left({\cal D}_{0+}^{1/2}g\right)(t-s)ds&=&\int_0^t\left({\cal D}_{0+}^{1/2}f\right)(s)g(t-s)ds\nonumber\\
&+&\left({\cal J}_{0+}^{1/2}f\right)(0)g(t)-f(t)\left({\cal J}_{0+}^{1/2}g\right)(0),
\end{eqnarray}
where by $\left({\cal J}_{0+}^{1/2}f\right)(0)$  is undewrstood the  $\lim_{\tau\rightarrow 0+}{\left({\cal J}^{1/2}_{0+}f\right) (\tau)}$,  if it exists.\\
For the case concerning the present work, Dargush and Kim  \cite{DARGUSH-et-al2012}  have reached at the  the interesting conclusion
$$
\int_0^t\left({\cal D}_{0+}^{1/2}u\right)(s)\left({\cal D}_{0+}^{1/2}u\right)(t-s)ds=\int_0^tu'(s)u(t-s)ds+u(0)u(t),
$$
or
\begin{equation}
\left[ {\cal D}_{0+}^{1/2}u,{\cal D}_{0+}^{1/2}u\right]=\left[ u',u\right]+u(0)u(t).
\end{equation}
 Hereafter, for the sake of simplicity  we will denote the left  half derivative as
\begin{equation}
{\cal D}^{1/2}_{0+}u(s)=u^{(1/2)}(s).
\end{equation}
Thus eq. (24)  is written
\begin{equation}
\left[u^{(1/2)},u^{(1/2)}\right]=\left[ u',u\right]+u(0)u(t).
\end{equation}
The final step to conclude this subsection is to present a formula for the first variation when a functional includes  half derivatives. 
Let  $u$ any differentiable function on $[0,t],\  t>0$ and  the functional
\begin{equation}
J[u]=\left[u^{(1/2)},u^{(1/2)} \right],\quad u(0)=u_0.
\end{equation}
The domain of $J$, that is, the  set of the admissible functions and the set of amissible variations are given by 
$$
{\cal D}=\{\phi \in C^1[0,t]\ |\ \phi(0)=u_0\}\quad\hbox{and}\quad {\cal D}_0=\{\phi \in C^1[0,t]\ |\ \phi(0)=0\},
$$
respectively. Thus,  any admissible function can be written as well
$$\tilde{u}=u+\epsilon\eta,$$ 
for $u\in {\cal D}$, and for any $\eta \in {\cal D}_0$ and   $\epsilon\in\Re$. Thus the variation of the functional  (27)  at $u$ can be calculated as
$$
D\left[u^{(1/2)},u^{(1/2)}\right](u;\eta )=\frac{d}{d\epsilon}J[\tilde{u}]|_{\epsilon=0}=2\left[u^{(1/2)},{\eta}^{(1/2)}\right], \quad\eta\in{\cal D}_0,
$$
which, on account of eq. (26), gives
$$
2\left[u^{(1/2)},{\eta}^{(1/2)}\right]=2\left([u',\eta] +u(0)\eta(t)\right),
$$
that is,
\begin{equation}
D\left[u^{(1/2)},u^{(1/2)}\right](u;\eta )=2\left([u',\eta] +u(0)\eta(t)\right).
\end{equation}
\\
In the case where $u=u(x,s)$, we will keep writing $u^{(1/2)}$, but it will be understood as half partial derivative with respect to $s$. 
With this convention, all above formulae will hold for functions of space and time. 
For instance, if $u$ is sufficiently smooth function of $x$ and $s$, eqs (26) and  (28) will take the form
\begin{equation}
\left[u^{(1/2)},u^{(1/2)}\right]=\left[ \frac{\partial u}{\partial s},u\right]+u(x,0)u(x,t).
\end{equation}
and 
\begin{equation}
D\left[u^{(1/2)},u^{(1/2)}\right](u;\eta )=2\left(\left[\frac{\partial u}{\partial s},\eta\right] +u(x,0)\eta(x,t)\right),
\end{equation}
respectively.
\subsection{A fundamental lemma}
In order to use the variational calculuus techniques, we need an appropriate version of the fundamental lemma in the calculus of variations. Thus we invoke the following  lemma that has been proved by Gurtin \cite{GURTIN1964}.\\
\\
{\bf Lemma 2.1} {\it Let $f$ a continuous function on $\bar{\Omega} \times [0,\infty )
$ and suppose that 
$$
\int_{\Omega}[f,v]d{\bf x}=0, \quad t<\infty
$$
for all $v\in C^{\infty}(\bar{\Omega}\times [0,t])$ that, together with all of their space derivatives, vanish on $\partial\Omega\times [0,t]$. Then
$$
f({\bf x},s)=0, \hbox{ for all }({\bf x},s)\in \bar{\Omega}\times [0,t].
$$
}
\section{Discrete Systems}
\subsection{Conservative systems}
To make the main point of this paper clear, we consider first the simplest dynamical problem: the motion of a conservative system of a single degree of fredom.
\subsubsection{A generalization of the standard Hamilton's principle}
Let the oscilation of a mass $m$,  attached to an ideal spring with constant $k$, on which applies a continuous force $f=f(s),\hspace{0.1 cm} s\in (0,\infty)$. The motion starts with an initial displacement $u_0$ and an initial velocity $v_0$.
 Denoting by $u(t)$ the displacement from the equilibrium position  of the mass at time $t$, the strong form of the initial value problem can be written\\
\\
{\bf Problem 3.1} {\it  Find a function $u\in C^2[0,t]$ which fulfills the following differential equation and initial conditions}
\begin{equation}
mu''(s)+ku(s)=f(s),  s\in(0,t],\quad 0<t<\infty,
\end{equation}
\begin{equation}
u(0)=u_0,  \  u'(0)=v_0.
\end{equation}
\\
As many authors have noted, the standard Hamilton's principle can reproduce succesfully the field equation (31) by the use of the following action functional
\begin{equation}
I[u]:=\int_0^t\left(\frac{m}{2}u'(s)u'(s)-\frac{k}{2}u(s)u(s)\right)ds-\int_0^tf(s)u(s)ds
\end{equation}
only if the value of $u$ at the end point $t$ is known, that is, only if $u$ meets the additional condition
 \begin{equation}
u(t)=u_1,
\end{equation}
where $u_1$ is given, i.e, an additional datum must be provided to the problem (31)-(32).  Actually the set of admmisssible functions accompanying the functional (33) becomes
\begin{equation}
{\cal D}=\{\phi\in C^2[0,t]\ | \ \phi(0)=u_0 \hbox{ and }  \phi(t)=u_1 \},
\end{equation}
Thus the set of variations will be of the form
\begin{equation}
{\cal D}_0=\{\phi\in C^2[0,t]\ | \ \phi(0)=0 \hbox{ and }  \phi(t)=0 \},
\end{equation}
Apart from that, in the procedure of derivation of eq. (31) there is no need for using the second initial condition (32b).\\
\\
In this subsection, we will make an attempt to formulate a plain, i.e, not a mixed, Lagrangian formulation which might be regarded as a direct generalization of the standard Hamilton's principle based on the action functional (33). 
Indeed, instead of eq. (33), we propose the definition of the action integral as follows
\begin{eqnarray}
I[u]&=&\frac{m}{2}\left[ u',u'\right]+\frac{k}{2}\left[ u,u\right]-[f,u]-[\tilde{f},u]\nonumber\\
&=&\int_0^t\left(\frac{m}{2}u'(s)u'(t-s)+\frac{k}{2}u(s)u(t-s)\right)ds\nonumber\\
&-&\int_{0^-}^t\left(f(s)u(s)+\tilde{f}(s)u (s)\right)ds,
\end{eqnarray}
with  
\begin{equation}
\tilde{f}(s)=f_0\delta (s),
\end{equation}
where $f_0$ is a constant and  $\delta$ is the Dirac function. Thus $\tilde{f}$ is an additional force that acts instantaneously at the initial time $s=0$. Later, the inrerpretation of this force will be further discussed.\\
\\
The domain of the functional (37), that is, the set of admissible functions will be
\begin{equation}
{\cal D}=\{\phi \in C^2 [0,t]\  | \ \phi(0)=u_0\}, 
\end{equation}
while the set of admissible variations is
\begin{equation}
{\cal D}_0=\{\phi \in C^2 [0,t]\ | \ \phi(0)=0\}.
\end{equation}
\\
{\bf Remark 3.1} The standard action integral (eq. (33)) consists of  two parts: the first one is related with the  difference between the kinetic and potential (elastic) energy and the second with the the work of the external force. 
In the  functional proposed here, one can find analogous parts but the kinetic like term is {\it  added} to the potential like term, i.e., it seems that the first part is related in some manner with the total energy of the system. 
This might be considered as an essential change of the action integral  stucture.\\
\\
Next, the following variational  principle is proved.
\subsubsection{A stationary principle for the functional (37)}
{\bf Proposition 3.1}
{\it If the functional (37) attains a stationary value at some $u\in {\cal D}$, then this  $u$ will  satisfy Problem 3.1, too.}\\
\\
{\bf Proof}: Suppose that  functional (37) attains a stationary value at  $u\in{\cal D}$. A necessary condition for that is the variation of $I$  at $u$ with respect to any $\eta\in{\cal D}_0$  to be zero.
\begin{equation}
DI(u;\eta)=\frac{d}{d\epsilon}I(u+\epsilon \eta){\mid}_{\epsilon =0}=0,\hbox{ for all }\eta\in{\cal D}_0.
\end{equation}

First, we calculate the variation of the last term of the action functional:
\begin{eqnarray}
\frac{d}{d\epsilon}[\tilde{f},\tilde{u}]|_{\epsilon=0}&=&\frac{d}{d\epsilon}\int_{0^-}^tf_0\delta(s)\left(u(t-s)+\epsilon\eta(t-s) \right)ds|_{\epsilon=0}\nonumber\\
&=&\int_{0^-}^tf_0\delta(s)\eta(t-s) ds=f_0\eta(t)
\end{eqnarray}
Using eqs. (15), (18) and (42), eq. (41) is calculated
\begin{eqnarray}
DI(u;\eta)&=&[mu'',\eta]+mu'(0)\eta(t)+k[u,\eta]-[f,\eta]-f_0\eta(t)\nonumber\\
&=&[mu''+ku-f,\eta]+(mu'(0)-f_0)\eta(t)=0, \forall\eta\in{\cal D}_0,
\end{eqnarray}
where, we  furthermore have accounted for the fact that $f$ is a fixed function.
Given that the variational equation (43) holds for any $\eta\in{\cal D}_0$, it will hold for any subset of ${\cal D}_0$, too. Thus, if  we select the subset
$$
C_0^2[0,t]=\{\phi\in C^2[0,])|\phi(0)=0\  \hbox{and}\  \phi(t)=0 \},
$$
eq. (43) will take the form
\begin{equation}
[mu''+ku-f,\eta]=0,\quad\hbox{for all}\quad\eta\in C_0^2[0,t]
\end{equation}
Accounting for eq. (5), we obtain
\begin{equation}
mu''(s)+ku(s)=f(s),\quad\hbox{for all }\quad s\in[0,t],
\end{equation}
that is, the field equation (31) has been recovered.
After eq. (45) has been obtained, eq. (43) becomes
$$
(mu'(0)-f_0)\eta(t)=0,\quad\hbox{for all}\quad\eta\in{\cal D}_0,
$$
which directly gives
\begin{equation}
f_0=mu'(0)=mv_0
\end{equation}
and the Proposition 3.1 has been proved.\\
\\
{\bf Remark 3.2} Notice that the second initial condition (32b) enters the picture with the aid of eq. (46), playing a role  analogous to that of the natural boundary condition in a boundary value problem. 
It is apparent that the proposed functional (37) is consistent with the initial value problem (31)-(32) as long as it derives the field euation (31), incorporates the intial conditions (32) without the need of the additional end condition (34).
Also, it is ineresting to note that the first initial condition, which is analogous to the essential boundary condition, finds itself in the set of admissible functions ${\cal D}$, whereas the second initial condition, which is 
a  Neumann--like condition, enters the last term of the functional (37).\\
\\
{\bf Remark 3.3} We note that the initial velocity is responsible for the rising of the singular force $\tilde{f}$. Looking for a possible physical  interpretation of that force, one may invoke the notion of the force impulse,
which is defined as the integral of a force over a specific time interval $[t_1,t_2]$. Consider the case of a particle's free motion, then the impulse of the applied force is the momentum difference $mv(t_2)-mv(t_1)$.  If the particle  posses
 a nonzero initial velocity $v_0$, i.e, an initial momentum, it needs a force which will rise its momentum from $0$ up to $mv_0$ instantaneously. In other words, it needs a force the instantaneous  impulse of which should equal $mv_0$.  
The standard applied force $f$ can not offer this because its instantantaneous impulse  by definition vanishes. Thus, in the case where the time inteval tends to zero, the use the delta function proves to be necessary.
\subsection{Dissipative systems}
Consider now a damped oscilator so as to add a dissipative force into the equation of motion. In that case the initial value problem takes the form\\
\\
{\bf Problem 3.2} {\it Find a function $u\in C^2(0,t)$ which fulfill the differential equation}

\begin{equation}
mu''(s)+cu'(s)+ku(s)=f(s),  s\in(0,t],\ 0< t<\infty,
\end{equation}
{\it and the initial conditions}
\begin{equation}
u(0)=u_0,   u'(0)=v_0,
\end{equation}
{\it where $c$ represents the damping coefficient.}\\
\\
Following the analysis of the previous subsection, we propose an action functional of the form
\begin{equation}
I[u]=\frac{m}{2}\left[ u',u'\right]+\frac{c}{2}\left[ u^{(1/2)}, u^{(1/2)}\right]+\frac{k}{2}\left[ u,u\right]-[f,u]-[\tilde{f},u],\quad u\in {\cal D},
\end{equation}
where $\tilde{f}$ is again of the form (38) and the set of admissible functions $\cal D$ is given again by eq. (39).

The following statement might be a generalization of the Hamilton's principle for the case of a dissipative system\\
\\
{\bf Proposition 3.2}
{\it If the functional (49) attains a stationary value at some $u\in {\cal D}$, then $u$ will  fulfill the initial value problem (47)-(48).}\\
\\
{\bf Proof} The proof of this statement is quite close to the proof of Proposition 1, thus we focus only on the second viscous term of the functional (49),
 which with the aid of eq. (28) becomes
\begin{equation}
D\left(\frac{c}{2}\left[u^{(1/2)},u^{(1/2)}\right]\right)(u;\eta )=[cu',\eta] +cu(0)\eta(t).
\end{equation}
After eq. (50), the variation of the functional (49) can easily be calculated as 
\begin{eqnarray}
DI(u;\eta)= \hspace{10.6cm}  \nonumber\\
=[mu'',\eta]+mu'(0)\eta(t)+[cu',\eta] +cu(0)\eta(t)+k[u,\eta]-[f,\eta]-f_0\eta(t)\nonumber\\
=[mu''+cu'+ku-f,\eta]+(mu'(0)+cu(0)-f_0)\eta(t)=0, \forall\eta\in{\cal D}_0,
\end{eqnarray}
With a reasoning similar to that of the previous proof, we conclude 
\begin{equation}
mu''(s)+cu'(s)+ku(s)=f(s), \hbox{ for all } s\in[0,t]
\end{equation}
and 
\begin{equation}
f_0=mu'(0)+cu(0)=mv_0+cu_0.
\end{equation}
Thus, the proof has been completed.\\
\\
{\bf Remark 3.4} We note that the viscous term contributes to the singular force $\tilde{f}$ through the initial displacement as does the inertial term through the initial velocity.  That is, a part of the momentum at the onset is provided by viscocity (due to non--zero  initial displacement).  Thus, we have an initial condition resembling the Robin type boundary conditions in boundary value problems.
In that case, $f_0$ might be vanished even for non-zero initial displacement and initial velocity.


\section{Continuous Systems}
In this section we focus our attention to  continuous bodies. First, we remind the classical Hamilton's principle of elastodynamics and comment its drawbacks. 
Then we  proceed to a new formulation of Hamilton's principle for a one dimensional elastic and viscoelastic body, based on temporal convolution.
\subsection{The standard Hamilton's principle for elastodynamics}
The probem of linear elastodynamics can be written as \\
\\
{\bf Problem 4.1} {\it  Find a $C^2$ function ${\bf u}={\bf u}({\bf x},s)$  and $C^1$ functions $\tau=\tau({\bf x},s)$, ${\bf e}={\bf e}({\bf x},s)$,  ${\bf x}\in \bar{\Omega}$, $s\in [0,t])$, $t>0$, which fulifill the equations}
\begin{eqnarray}
&&\frac{\partial \tau}{\partial {\bf x}}+{\bf f}=\rho\frac{\partial^2 {\bf u}}{\partial s^2},\\
&&\tau={\bf C}{\bf e}, \\
&&{\bf e}=\frac{1}{2}\left( \nabla {\bf u}+{\nabla {\bf u}}^T\right),
\end{eqnarray}
{\it the boundary conditions}
\begin{eqnarray}
&&{\bf u}({\bf x},s)=\hat{\bf u}({\bf x},s),\quad ({\bf x},s)\in \partial \Omega_D\times [0,t],\\
&&{\bf t}({\bf x},s)=\hat{\bf P}({\bf x},s),\quad ({\bf x},s)\in \partial \Omega_N\times [0,t]
\end{eqnarray}
{\it and the initial conditions}
\begin{eqnarray}
&&{\bf u}({\bf x},0)={\bf u}_0({\bf x}),\quad {\bf x}\in \bar{\Omega},\\
&&\frac{\partial{\bf u}}{\partial{\bf s}}({\bf x},0)={\bf v}_0({\bf x}),\quad {\bf x}\in \bar{\Omega},
\end{eqnarray}
{\it where  $\Omega$ is an open domain of $E^3$, $\bf u$ the displacement field, $\tau$ the Cauchy stress tensor, $\bf f$ is the volume forces, $\rho$ the mass density, $\bf C$ the elasticity tensor, 
$\bf e$ the infinitesimal strain tensor, ${\bf t}=\tau {\bf n}$ the stress vector, $\bf n$ the outward unit normal vector along the surface $\partial \Omega$, $\hat{\bf u}$, $\hat{\bf P}$, ${\bf u}_0$ and  ${\bf v}_0$ are given functions.}\\

Also, with $\partial\Omega_N$ and  $\partial\Omega_D$  are denoted the parts of the boundary where the Neumann and the Dirichlet conditions hold, respectively. Moreover, it holds that 
$$
\partial\Omega_D\cup\Omega_N=\partial\Omega,\quad \partial\Omega_D\cap\Omega_N=\emptyset.
$$
\\
According to the standard Hamilton's principle, the action functional is defined as 
\begin{eqnarray}
I[{\bf u}]=
\int_{\Omega}\left[\frac{\rho}{2}\left(\frac{\partial{\bf u}}{\partial s},\frac{\partial{\bf u}}{\partial s}\right)-\frac{1}{2}\left({\bf C} {\bf e},{\bf e}\right)\right]d{\bf x}-\hspace{4 cm}\nonumber\\
-\int_{\Omega}\left( {\bf f},{\bf u}\right)d{\bf x}-\int_{\partial\Omega}\left(\hat{ {\bf P}},{\bf u}\right)d{\bf s},\ {\bf  u}\in {\cal D},
\end{eqnarray}
where 
\begin{equation}
{\cal D}=\{{\bf w}\in\left(C^2(\bar{\Omega}\times [0,t]\right):{\bf w}({\cdot},0)={\bf u}_0,   {\bf w}({\cdot},t)={\bf u}_1 \hbox{ and } {\bf w}(\partial{\Omega}_D,s)=\hat{\bf w} \}
\end{equation}
and with $(\ ,\ )$ is denoted the $L^2$ innere product.\\
Notice that apart from the initial time $s=0$,  an additional condition, concerning the displacement field at the final time $s=t$ is included in ${\cal D}$. \\
The Hamilton's principle for linear elastodynamics can be stated as follows\\
\\
{\bf Proposition 4.1. Hamilton's principle for elastodynamics}\\
{\it Let the functions ${\bf e}, \tau \hbox{ and }{\bf u}\in{\cal D} $ satisfy eqs. (55)-(56).  If, in addition, the functional (59) attains a stationary value at the dispacement filed ${\bf u}$, then  $({\bf e}, \tau,{\bf u} )$ is a solution of Problem 4.1.}
\subsection{A generalized Hamilton's  principle for the motion of an elastic bar}
Based on concepts discused in the preceding sections, we present and  prove  a variational statement which might be viewed as a direct generalization of the Hamilton's principle.
For simplicity reasons, we are limited only in a one dimensional elastic body. The generalization to two or three dimensions is straightforward.\\
\\
Consider the motion along the axis of  a linear elastic bar constrained at its left end and loaded by a force distributed over its length and another one applied  at its right end. 
The equation of motion, the constitutive  and the kinematic relations are written
\begin{eqnarray}
&&\frac{\partial\tau}{\partial x}+f=\rho\frac{\partial^2u}{\partial s^2},\quad (x,s)\in (0,l)\times (0,t],\\
&&\tau=Ee,\quad e=\frac{\partial u}{\partial x},
\end{eqnarray}
where $u=u(x,s)$ is the displacement field, $\tau=\tau(x,s)$ is the stress field, $e=e(x,s)$ is the linear strain, $f=f(x,s)$ is the distributed force per unit length, $E$ the Young modulus and $\rho$ the mass density per unit length.

Combining the three above equations one obtains the field equation in terms of displacement:
\begin{equation}
\rho\frac{\partial^2u}{\partial s^2}-E\frac{\partial^2u}{\partial x^2}=f,\quad(x,s)\in(0,l)\times(0,t].
\end{equation}
Thus, the final formulation of the complete initial--boundary value problem can be written\\
\\
{\bf Problem 4.2} {\it Find a sufficiently smooth function $u=u(x,s)$ which fulfills the field equation (65) as well as the following initial and boundary conditions\\
\\
\begin{eqnarray}
&&u(x,0)=u_0(x), \quad x\in [0,l],\\
&&\frac{\partial u}{\partial s}(x,0)=v_0(x),\quad x\in[0,l]
\end{eqnarray}
 and
\begin{eqnarray}
&&u(0,s)=\hat{u}(s), \quad s\in [0,t],\\
&&E\frac{\partial u}{\partial x}(l,s)=p(s),\quad s\in[0,t].
\end{eqnarray}
respectively, where $u_0$, $v_0$, $\hat{u}$, $f$ and $p$ are given continuous functions, with $u_0(0)=\hat{u}(0)$.}\\
\\
We will show that the above problem admits a variational formulation of Hamilton's type. 
We start with the definition of the action functional
\begin{eqnarray}
J[u]=
\int_0^l\left(\frac{\rho}{2}\left[\frac{\partial u}{\partial s},\frac{\partial u}{\partial s}\right]+\frac{E}{2}\left[\frac{\partial u}{\partial x},\frac{\partial u}{\partial x}\right]\right)dx-\hspace{2cm}\nonumber\\
-\int_0^l\left([f,u]-[\hat{f},u]\right)dx-\left[p,u(l,\cdot)\right], \ u\in{\cal D},
\end{eqnarray}
where
\begin{equation}
{\cal D}=\{\phi\in C^2\left([0,l]\times[0,t]\right)\ |\ \phi(0,\cdot)= \hat{u}\hbox{ and }\phi(\cdot,0)=u_0\}
\end{equation}
and
\begin{equation}
\hat{f}(x,s)=\hat{f}_0(x)\delta(s).
\end{equation}
\\
{\bf Remark 4.1} Notice that an extra term, i.e., $[\hat{f},u]$ contributes to what might be called "the action of the work of the external forces". It is a "new" volume force acting instantaneously at time $s=0$.
It will be shown that this force is responsible for the initial velocity $v_0(x)$. It is worth noting that the presence of datum for the time derivative of $u$ at $s=0$ gives rise to the presence of the volume force, $\hat{f}$ in the action functional, like the presence of datum for the spatial derivative of $u$ at $x=l$ causes the presence of the traction, $p$ in the action functional. \\
\\
The set of variations that corresponds to eq. (71) will be 
\begin{equation}
{\cal D}_0=\{\phi\in C^2\left([0,l]\times[0,t]\right)\ |\ \phi(0,s)= 0=\phi(x,0)\}.
\end{equation}
Then, one can prove the following  variational statement\\
\\
{\bf Proposition 4.2} {\it If the functional $J$ attains a stationary value at $u\in{\cal D}$, then this $u$ solves the Problem 4.2.}\\
\\
{\bf Proof} Let a function  $u\in{\cal D}$ at which $J$ takes a stationary value. A necessary condtion for that is 
\begin{equation}
DJ(u;v)=\frac{d}{d\epsilon}J[u+\epsilon v]|_{\epsilon =0}=0 , \hbox{ for all } v\in{\cal D}_0.
\end{equation}
We calculate the first term of the variation of $J$:
\begin{eqnarray}
&&D\left(\int_0^l\frac{\rho}{2}\left[\frac{\partial u}{\partial s},\frac{\partial u}{\partial s}\right]dx \right)(u;v)=\int_0^l\rho\left[\frac{\partial u}{\partial s},\frac{\partial v}{\partial s}\right]dx \nonumber\\
&&=\int_0^l\int_0^t\rho\frac{\partial u}{\partial s}(x,s)\frac{\partial v}{\partial z}(x,z)dsdx,\quad z=t-s\nonumber\\
&&=-\int_0^l\int_0^t\rho\frac{\partial u}{\partial s}(x,s)\frac{\partial v}{\partial s}(x,t-s)dsdx,\nonumber\\
&&=-\int_0^l\left(\left[\rho\frac{\partial u}{\partial s}(x,s)v(x,t-s)\right]_0^t -\int_0^t\frac{\partial^2 u}{\partial s^2}(x,s)v(x,t-s)ds \right)dx\nonumber\\
&&=-\int_0^l\left(-\rho\frac{\partial u}{\partial s}(x,0)v(x,t)-\int_0^t\frac{\partial^2 u}{\partial s^2}(x,s)v(x,t-s)ds \right)dx\nonumber\\
&&=\int_0^l\left[\rho\frac{\partial^2 u}{\partial s^2},v\right]dx+\int_0^l\rho\frac{\partial u}{\partial s}(x,0)v(x,t)dx.
\end{eqnarray}
Similarly, one can calculate the second term of $DJ(u;v)$:
\begin{eqnarray}
&&D\left(\int_0^l\frac{E}{2}\left[\frac{\partial u}{\partial x},\frac{\partial u}{\partial x}\right]dx \right)(u;v)=\nonumber\\
&&-\int_0^l\left[E\frac{\partial^2 u}{\partial x^2},v\right]dx+\int_0^tE\frac{\partial u}{\partial x}(l,s)v(l,t-s)ds.
\end{eqnarray}
The variation of the next term will be of the form
\begin{eqnarray}
&&D\left(\int_0^l([f,u]+[\hat{f},u])dx\right)(u;v)=\int_0^l([f,v]+[\hat{f},v])dx\nonumber\\
&&=\int_0^l\left([f,v]+\int_{0^-}^t\hat{f}(x,s)v(x,t-s)ds\right)dx\nonumber\\
&&=\int_0^l\left([f,v]+\int_{0^-}^t\hat{f}_0(x)\delta(s)v(x,t-s)ds\right)dx\nonumber\\
&&=\int_0^l[f,v]dx+\int_0^l\hat{f}_0(x)v(x,t)dx.
\end{eqnarray}
What remains is the last term 
\begin{eqnarray}
D\left([p,u(l,\cdot)]\right)(u;v)=[p,v(l,\cdot)]=\int_0^tp(s),v(l,t-s)ds.
\end{eqnarray}
Inserting  eqs (75-78) into eq (74), one obtains the variation of $J$
\begin{eqnarray}
DJ(u;v)=\hspace{9.5cm}\nonumber\\
\int_0^l\left[ \rho\frac{\partial^2 u}{\partial s^2}-E\frac{\partial^2 u}{\partial x^2}-f,v\right]dx+\int_0^l\left(\rho\frac{\partial u}{\partial s}(x,0)-\hat{f}_0(x)\right)v(x,t)dx\nonumber\\
+\int_0^t\left(E\frac{\partial u}{\partial x}(l,s)-p(s)\right)v(l,t-s)ds=0, \hbox{ for all } v \in {\cal D}_0
\end{eqnarray}
If we confine ourselves in the subset of ${\cal D}_0$:
\begin{equation}
C_0^2\left([0,l]\times[0,t] \right)=\{\phi\in {\cal D}_0|\phi(l,s)=0 \hbox{ and }\phi(x,t)=0\},
\end{equation}
we will get
\begin{equation}
\int_0^l\left[ \rho\frac{\partial^2 u}{\partial s^2}-E\frac{\partial^2 u}{\partial x^2}-f,v\right]dx=0, \quad\hbox{for all } v \in C_0^2\left([0,l]\times[0,t] \right).
\end{equation}
Recalling Lemma 2.1, we  obtain
\begin{equation}
 \rho\frac{\partial^2 u}{\partial s^2}-E\frac{\partial^2 u}{\partial x^2}-f=0, \hbox{ for all }(x,s)\in [0,l]\times[0,t],
\end{equation}
thus the field equation (65) has been recovered.\\
Given that eq. (82) holds, the variational equation (79) becomes
\begin{eqnarray}
\int_0^l\left(\rho\frac{\partial u}{\partial s}(x,0)-\hat{f}_0(x)\right)v(x,t)dx\hspace{6cm}\nonumber\\
+\int_0^t\left(E\frac{\partial u}{\partial x}(l,s)-p(s)\right)v(l,t-s)ds=0, \hbox{ for all } v \in {\cal D}_0.
\end{eqnarray}
Confining ourselves in that subset of ${\cal D}_0$ where  $v(l,s)=0,$ for all $s\in[0,t]$, we obtain
\begin{equation}
\int_0^l\left(\rho\frac{\partial u}{\partial s}(x,0)-\hat{f}_0(x)\right)v(x,t)dx=0,\hbox{ for all } v(\cdot,t) \in C_0^2[0,l],
\end{equation}
which by the standard lemma of variational calculus  \cite{BEDFORD} provides
\begin{equation}
\rho v_0(x)=\hat{f}_0(x),\hbox{ for all }x\in[0,l],
\end{equation}
where the initial condition (67) has been used.\\
Returning to eq. (83) and making the selection $v(x,t)=0$  for all $x\in(0,l)$, one obtains
\begin{equation}
\int_0^t\left(E\frac{\partial u}{\partial x}(l,s)-p(s)\right)v(l,t-s)ds=0,\hbox{ for all } v(l,\cdot) \in C_0^2[0,t].
\end{equation}
The latter equation, with the aid of eq. (5),  gives
\begin{equation}
E\frac{\partial u}{\partial x}(l,s)=p(s),\hbox{ for all } x \in [0,t],
\end{equation}
which is the boundary condition (69), thus the Proposition 4.2  has been proved.\\

\subsection{A stationary principle for a linear viscoelastic bar}
Consider now the uniaxial motion of a bar characterized by a viscoelastic response of Kelvin-Voigt type
\begin{equation}
\tau(x,s)=Ee(x,s)+\gamma \frac{\partial e}{\partial s}(x,s), \quad (x,s) \in [0,l])\times [0,t],
\end{equation}
where $E$ and $\gamma$ are the elasticity and viscocity coefficients, respectively.\\
Inserting the consitutive relation (eq. (88)) into the equaton of motion, (63),  and using the kinematic relation, eq. (64b), one obtains
\begin{equation}
E\frac{\partial^2u}{\partial x^2}+\gamma\frac{\partial^3u}{\partial x^2\partial s}+f=\rho \frac{\partial^2u}{\partial s^2},\quad (x,s) \in (0,l)\times (0,t).
\end{equation}
The  initial--boundary value problem for a viscoelastic bar, formulated with respect to the displacements field takes the form\\
\\
{\bf Problem 4.3} {\it Find a sufficiently smooth function $u=u(x,s)$ which fulfills the field equation (89) as well as the following   initial and boundary conditions
\begin{eqnarray}
&&u(x,0)=u_0(x), \quad x\in [0,l],\\
&&\frac{\partial u}{\partial s}(x,0)=v_0(x),\quad x\in[0,l],\\
&&u(0,s)=\hat{u}(s), \quad s\in [0,t],\\
&&E\frac{\partial u}{\partial x}(l,s)+\gamma \frac{\partial^2u}{\partial x\partial s}(l,s)=p(s),\quad s\in[0,t],\\
&&\gamma \frac{\partial u}{\partial x}(l,0)=\hat{p}_0,
\end{eqnarray}
where $u_0$, $v_0$, $\hat{u}$,  $p$ and $f$ are given continuous functions with $u_0(0)=\hat{u}(0)$ and $\hat{p}_0$ is given constant.}\\
\\
We will show that the above problem admits a variational formulation. To this end,
we define the action functional 
\begin{eqnarray}
J[u]=
\int_0^l\left(\frac{\rho}{2}\left[\frac{\partial u}{\partial s},\frac{\partial u}{\partial s}\right]+\frac{E}{2}\left[\frac{\partial u}{\partial x},\frac{\partial u}{\partial x}\right]
+\frac{\gamma}{2}\left[\frac{\partial u}{\partial x}^{(1/2)},\frac{\partial u}{\partial x}^{(1/2)}  \right]\right)dx-\nonumber \\
-\int_0^l\left([f+\hat{f},u]\right)dx
-\left[p+\hat{p},u(l,\cdot)\right],  u\in{\cal D},
\end{eqnarray}
where
\begin{equation}
{\cal D}=\{\phi\in C^3\left([0,l]\times[0,t]\right)|\phi(0,\cdot)= \hat{u}\hbox{ and }\phi(\cdot ,0)=u_0\}
\end{equation}
and
\begin{equation}
\hat{f}(x,s)=\hat{f}_0(x)\delta(s)\quad \hat{p}(s)=\hat{p}_0\delta(s).
\end{equation}\\
{\bf Remark 4.2} In continuation to the Remark 4.1, $\hat{f}$ is singular volume force acting only at the initial moment, $\hat{f}_0$ is a fixed continuous function which will be proved that represents the total initial momentum of the body provided by the initial velocity and  initial strain.  By analogy, the  instantaneous contact force $\hat{p}$ has been intoduced to match with the condition (94).\\
\\
The set of variations that correspond to eq. (96) will be 
\begin{equation}
{{\cal D}_0}=\{\phi\in C^3\left([0,l]\times[0,t]\right)|\phi(0,s)= 0=\phi(x,0)\}.
\end{equation}
Then, one can prove the following  variational statement\\
\\
{\bf Proposition 4.3} {\it If the functional $J$ attains a stationary value at $u\in{\cal D}$, then this $u$ solves the Problem 4.3.}\\
\\
{\bf Proof} Supose that the action functional obtains a stationary value at some $u\in{\cal D}$, then the variation of $J$ at $u$ with respect to any $v\in {\cal D}_0$ should vanish.
\begin{equation}
DJ(u;v)=\frac{d}{d\epsilon}J[u+\epsilon v]_{\epsilon=0},\hbox{ for all }v\in {\cal D}_0.
\end{equation}
\\
We focus only on the extra terms of functional (95)  in comparison with the ones of  functional (70) which we have already calculated in the last subsection. That is, we will negotiate the third and last term of the functional (95).\\
The variation of the viscous term of (95) is calcualted as follows
$$
D\left(\int_0^l\frac{\gamma}{2}\left[\frac{\partial u}{\partial x}^{(1/2)},\frac{\partial u}{\partial x}^{(1/2)} \right]dx \right)(u;v)=\int_0^l\gamma\left[\frac{\partial u}{\partial x}^{(1/2)},\frac{\partial v}{\partial x}^{(1/2)}\right]dx
$$
The latter with the aid of eq. (29) can be written
\begin{eqnarray}
\int_0^l\gamma\left[\frac{\partial u}{\partial x}^{(1/2)},\frac{\partial v}{\partial x}^{(1/2)} \right]dx=\int_0^l\gamma\left[\frac{\partial}{\partial s}\left(\frac{\partial u}{\partial x}\right),\frac{\partial v}{\partial x} \right]dx\nonumber\\
+\int_0^l\gamma\left(\frac{\partial u}{\partial x}(x,0)\frac{\partial v}{\partial x}(x,t)
-\frac{\partial u}{\partial x}(x,t)\frac{\partial v}{\partial x}(x,0)\right)dx
\end{eqnarray}
In the sequel, we calculate eq. (100) term by term 
\begin{eqnarray}
\bullet&&\int_0^l\gamma\left[\frac{\partial^2 u}{\partial x \partial s},\frac{\partial v}{\partial x} \right]dx=\int_0^t\int_0^l\gamma\frac{\partial^2 u}{\partial x \partial s}(x,s)\frac{\partial v}{\partial x} (x,t-s)dxds\nonumber\\
&&=\int_0^t\left(\left[\gamma\frac{\partial^2 u}{\partial x \partial s}(x,s)v (x,t-s)\right]_0^l- \int_0^l\gamma\frac{\partial^3 u}{\partial x^2 \partial s}(x,s) v(x,t-s)dx\right)ds\nonumber\\
&&=\int_0^t\gamma\frac{\partial^2 u}{\partial x \partial s}(l,s)v (l,t-s)ds-\int_0^t\int_0^l\gamma\frac{\partial^3 u}{\partial x^2 \partial s}(x,s)v (x,t-s)dxds\nonumber\\
&&=-\int_0^l\left[\gamma\frac{\partial^3 u}{\partial x^2 \partial s},v\right]dx+\int_0^t\gamma\frac{\partial^2 u}{\partial x \partial s}(l,s)v (l,t-s)ds.\\
\bullet&&\int_0^l\gamma\frac{\partial u}{\partial x}(x,0)\frac{\partial v}{\partial x}(x,t)dx=\left[\gamma\frac{\partial u}{\partial x}(x,0)v(x,t)\right]_0^l-\int_0^l\gamma\frac{\partial^2 u}{\partial x^2}(x,0)v(x,t)dx\nonumber\\
&&=-\int_0^l\gamma\frac{\partial^2 u}{\partial x^2}(x,0)v(x,t)dx+\gamma\frac{\partial u}{\partial x}(l,0)v(l,t).\\
\bullet&&\int_0^l\gamma\frac{\partial u}{\partial x}(x,t)\frac{\partial v}{\partial x}(x,0)dx=0.
\end{eqnarray}
Finally, after eqs (100--103),  the variation of the viscous term can be written
\begin{eqnarray}
D\left(\int_0^l\frac{\gamma}{2}\left[\frac{\partial u}{\partial x}^{(1/2)},\frac{\partial u}{\partial x}^{(1/2)} \right]dx \right)(u;v)=\nonumber\\
=-\int_0^l\left[\gamma\frac{\partial^3 u}{\partial x^2 \partial s},v\right]dx+\int_0^t\gamma\frac{\partial^2 u}{\partial x \partial s}(l,s)v (l,t-s)ds\nonumber\\
-\int_0^l\gamma\frac{\partial^2 u}{\partial x^2}(x,0)v(x,t)dx+\gamma\frac{\partial u}{\partial x}(l,0)v(l,t).
\end{eqnarray}
\\
The variation of the last term of functional (95)  is easily calculated 
\begin{eqnarray}
D\left([p+\hat{p},u(l,\cdot)]\right)(u;v)=\int_0^t p(s)v(l,t-s)ds+\int_{0-}^t \hat{p}_0\delta (s)v(l,t-s)ds\nonumber\\
=\int_0^t p(s)v(l,t-s)ds+\hat{p}_0v(l,t).
\end{eqnarray}

Inserting (75--78) and (104--105) into eq. (99), one obtains
\begin{eqnarray}
DJ(u;v)&=&\int_0^l\left[ \rho\frac{\partial^2 u}{\partial s^2}-E\frac{\partial^2 u}{\partial x^2}-\gamma\frac{\partial^3 u}{\partial x^2 \partial s}-f,v\right]dx\nonumber\\
&&+\int_0^l\left(\rho\frac{\partial u}{\partial s}(x,0)-\gamma\frac{\partial^2 u}{\partial x^2}(x,0)-\hat{f}_0(x)\right)v(x,t)dx\nonumber\\
&&+\int_0^t\left(E\frac{\partial u}{\partial x}(l,s)+\gamma\frac{\partial^2 u}{\partial x \partial s}(l,s)-p(s)\right)v(l,t-s)ds\nonumber\\
&&+\left(\gamma\frac{\partial u}{\partial x}(l,0)-\hat{p}_0\right)v(l,t)=0, \hbox{ for all } v \in {\cal D}_0.
\end{eqnarray}
Confining the set of admissible variations only to those ones which vanish on the straight lines  $x=l$ and $s=t$, i.e., those $v$ that moreover  satisfy:
$$
v(l,s)=0, \hbox{ for all } s\in[0,t]
$$ 
 and 
$$
v(x,t)=0, \hbox{ for all } x\in[0,l],
$$
the above  variational equation (eq. (106)) becomes
\begin{equation}
\int_0^l\left[ \rho\frac{\partial^2 u}{\partial s^2}-E\frac{\partial^2 u}{\partial x^2}-\gamma\frac{\partial^3 u}{\partial x^2 \partial s}-f,v\right]dx=0,\hbox{ for all } v \in C^3_0([0,l]\times [0,t]).
\end{equation}
Thus, using Lemma 2.1, one obtains
\begin{equation}
 \rho\frac{\partial^2 u}{\partial s^2}-E\frac{\partial^2 u}{\partial x^2}-\gamma\frac{\partial^3 u}{\partial x^2 \partial s}-f=0,\hbox{ for all } (x,s)\in [0,l]\times [0,t],
\end{equation}
i.e., the field equation (89).\\
Returning to eq. (106) and acconting for eq. (108), we can write
\begin{eqnarray}
&&\int_0^l\left(\rho\frac{\partial u}{\partial s}(x,0)-\gamma\frac{\partial^2 u}{\partial x^2}(x,0)-\hat{f}_0(x)\right)v(x,t)dx\nonumber\\
&&+\int_0^t\left(E\frac{\partial u}{\partial x}(l,s)+\gamma\frac{\partial^2 u}{\partial x \partial s}(l,s)-p(s)\right)v(l,t-s)ds\nonumber\\
&&+\left(\gamma\frac{\partial u}{\partial x}-\hat{p}_0\right)(l,0)v(l,t)=0, \hbox{ for all } v \in {\cal D}_0.
\end{eqnarray}
Restricting the set  ${\cal D}_0$ only to those variations which vanish on $x=l$, the variational equation (109) can be written
$$
\int_0^l\left(\rho\frac{\partial u}{\partial s}(x,0)-\gamma\frac{\partial^2 u}{\partial x^2}(x,0)-\hat{f}_0(x)\right)v(x,t)dx=0, \hbox{ for all } v(\cdot ,t) \in C_0^3[0,l],
$$
from which one, invoking the standard lemma of variational calculus, obtains
\begin{eqnarray}
\hat{f}_0(x)=\rho v_0(x)-\gamma\frac{\partial^2 u}{\partial x^2}(x,0), \hbox{ for all } x \in [0,l].
\end{eqnarray}
Thus, eq. (109) is written
\begin{eqnarray}
\int_0^t\left(E\frac{\partial u}{\partial x}(l,s)+\gamma\frac{\partial^2 u}{\partial x \partial s}(l,s)-p(s)\right)v(l,t-s)ds\nonumber\\
+\gamma\frac{\partial u}{\partial x}(l,0)v(l,t)=0, \hbox{ for all } v \in {\cal D}_0,
\end{eqnarray}
which for $v(l,t)=0$, gives rise to 
\begin{eqnarray}
\int_0^t\left(E\frac{\partial u}{\partial x}(l,s)+\gamma\frac{\partial^2 u}{\partial x \partial s}(l,s)-p(s)\right)v(l,t-s)ds=0,\hspace{1 cm}\nonumber\\ \hbox{for all}\quad   v(l,\cdot) \in C_0^3[0,t].
\end{eqnarray}
Thus invoking eq. (5), one can take 
\begin{equation}
E\frac{\partial u}{\partial x}(l,s)+\gamma\frac{\partial^2 u}{\partial x \partial s}(l,s)=p(s),  \hbox{ for all } s \in [0,t],
\end{equation}
which is nothing but the initial condition (93).\\
What remains from the variational equation is a simple product of real numbers:
\begin{equation}
\left(\gamma\frac{\partial u}{\partial x}(l,0)-\hat{p}_0\right)v(l,t)=0, \hbox{ where } v(l,t)   \hbox{ is any real, }
\end{equation}
thus one concludes
\begin{equation}
\gamma\frac{\partial u}{\partial x}(l,0)=\hat{p}_0,
\end{equation}
which is the condition (94), thus the proof of Proposition 4.3 has been completed.\\
\section{Conclusions}
Using the notion of convolution, the fractional derivative of order $1/2$ and the Dirac function,  we have  provided stationary principles for dynamics of conservative and dissipative systems. The action functional as well as the Euler--Lagrange equations were formulated in terms of the displacement field which, in all cases, is the sole unknown variable of the variational problem. \\
\\
We examined the forced oscilation for both damped and undamped cases as examples of Hamilton principles for discrete systems. As regards the continuous media, the total potential energy principle has been consistenly generalized to Hamilton principle for one--dimensional elastic and viscoelastic body. In all cases, we arrived at variational principles which harmoniously  incorporate all the initial conditionns and systematically account for dissipation. \\
\\
It is worth noting that the initial and boundary conditions contribute evenly to the "construction" of the action functional: the natural conditions provide, as usual,  a term through the tractions, while the initial velocity provides an additional term through a volume force of Dirac type. Also, initial and boundary conditions contribute together in the definition of the kinematically admissible functions.  From this point of view, one can claim that the proposed Hamilton's principle is a consistent generalization to dynamics of the total potential energy principle.\\
\bibliographystyle{plain}
\bibliography{Hamilton}
\end{document}